\documentclass[aps, prd, twocolumn, amsmath, floats,floatfix, superscriptaddress, nofootinbib,
showpacs]{revtex4}

\usepackage{amssymb}
\usepackage{amsmath}
\usepackage{verbatim}
\usepackage{mathrsfs}
\usepackage{amsfonts}
\usepackage{latexsym}
\usepackage{epsfig}
\usepackage{color}
\usepackage{graphicx,subfigure}
\usepackage{units}
\usepackage{envmath}
\usepackage{natbib}
\definecolor{verd}{rgb}{0.13, 0.55, 0.13}

\begin{document}

\title{Quasistationary solutions of scalar fields around collapsing self-interacting boson stars}
\author{Alejandro Escorihuela-Tom\`as}
\affiliation{Departamento de
  Astronom\'{\i}a y Astrof\'{\i}sica, Universitat de Val\`encia,
  Dr. Moliner 50, 46100, Burjassot (Val\`encia), Spain}
\author{Nicolas Sanchis-Gual}
\affiliation{Departamento de
  Astronom\'{\i}a y Astrof\'{\i}sica, Universitat de Val\`encia,
  Dr. Moliner 50, 46100, Burjassot (Val\`encia), Spain}

\author{Juan Carlos Degollado} 
\affiliation{
Instituto de Ciencias F\'isicas, Universidad Nacional Aut\'onoma de M\'exico,
Apdo. Postal 48-3, 62251, Cuernavaca, Morelos, M\'exico.}

\author{Jos\'e A. Font}
\affiliation{Departamento de
  Astronom\'{\i}a y Astrof\'{\i}sica, Universitat de Val\`encia,
  Dr. Moliner 50, 46100, Burjassot (Val\`encia), Spain}
\affiliation{Observatori Astron\`omic, Universitat de Val\`encia, C/ Catedr\'atico 
  Jos\'e Beltr\'an 2, 46980, Paterna (Val\`encia), Spain}


\date{\today}


\begin{abstract}  
There is increasing numerical evidence that scalar fields can form 
long-lived quasi-bound states around black holes. Recent perturbative and 
numerical relativity calculations have provided further confirmation in a 
variety of physical systems, including both static and accreting black holes, 
and collapsing fermionic stars. In this work we investigate this issue yet again 
in the context of gravitationally unstable boson stars leading to black hole 
formation. We build a large sample of spherically symmetric initial models, both 
stable and unstable, incorporating a self-interaction potential with a quartic 
term. The three different outcomes of unstable models, namely migration to the 
stable branch, total dispersion, and collapse to a black hole,  are also present 
for self-interacting boson stars. Our simulations show that for 
black-hole-forming models, a scalar-field remnant is found outside the black-hole 
horizon, oscillating at a different frequency than that of the original 
boson star. This result is in good agreement with recent spherically symmetric 
simulations of unstable Proca stars collapsing to black 
holes~\cite{sanchis2017numerical}.
\end{abstract}




\maketitle

\section{Introduction}\label{sec:introduction}


The development of stable numerical relativity codes based on hyperbolic formulations
of Einstein's equations, accompanied with suitable gauge conditions, has been 
critical for recent advances in our understanding of astrophysical systems 
involving strong gravity. In particular, those technical developments have allowed accurate 
numerical evolutions of highly dynamical spacetimes up to, and well beyond, the formation 
of black holes. Further steps have also been taken with the incorporation of matter 
content in black-hole spacetimes, specifically in the form of scalar fields, a type 
of matter that has found recurrent use in numerical relativity. Within such a 
framework, recent studies of the Einstein-Klein-Gordon (EKG) system, both in 
the linear and nonlinear regime, have shown that massive scalar fields surrounding 
black holes can accommodate a type of oscillatory mode which only decays at 
infinity~\cite{Witek:2012tr,Burt:2011pv,Barranco:2012qs,Barranco:2013rua,
sanchis2015quasistationary2,sanchis2015quasistationary,sanchis2016}. These 
quasi-bound states may thus linger around the black hole in the form of a long-lived 
remnant (a \emph{wig}) of scalar field. For both, scalar fields around supermassive 
black holes and axion-like scalar fields  around  primordial  black holes,  it  has  
been  found  that the  fields can indeed survive for cosmological timescales~\cite{Barranco:2012qs}.
Moreover, for spinning black holes, quasi-bound states can yield exponentially
growing modes~\cite{Detweiler:1980uk,Dolan:2007mj} and hairy-black-hole 
solutions~\cite{Herdeiro:2014goa}.


On the other hand, scalar fields are also known to allow for soliton-like 
solutions, i.e.~static,  spherically  symmetric  solutions  of  the  EKG system 
for a massive and  complex field~\cite{Kaup:1968zz,Ruffini:1969qy},  which  are  
commonly known  as boson stars (see~\cite{Liebling:2012fv} for a review). The 
dynamical fate of boson stars has been thoroughly investigated numerically, both 
using perturbation theory~\cite{Gleiser:1988ih,Lee:1988av} and fully nonlinear 
numerical 
simulations~\cite{Seidel:1990jh,Balakrishna:1997ej,Hawley:2000dt,Guzman:2004jw}. 
Ref.~\cite{Seidel:1990jh} in particular first showed that the fate of unstable 
boson-star solutions was either the formation of a black hole or the migration of the star 
to the stable branch, regardless of the sign of the binding energy. A third outcome for 
unstable boson stars is their total 
dispersion~\cite{Balakrishna:1997ej,Guzman:2004jw,guzman2009three} a situation 
which only happens for boson stars with negative binding energy.


In this work we build a comprehensive sample of initial models of boson stars, 
incorporating a self-interaction potential with a quartic term. The inclusion of 
such self-interaction provides extra pressure support against gravitational 
collapse and increases the range of possible maximum masses of boson stars, 
allowing to encompass models with astrophysical significance. Here, we  
revisit the stability of the solutions for different values of the 
self-interaction coupling constant $\lambda$, incorporating values as large as 
$\Lambda\equiv\lambda/4\pi G \mu^2=100$, not previously accounted for (here 
$\mu$ is the bare mass of the scalar field). Our findings are consistent with 
the three different outcomes for unstable models, namely migration to the stable 
branch, total dispersion, and collapse to a black hole, reported previously for both the 
$\lambda=0$ (mini-)boson star case and for self-interacting boson 
stars (see~\cite{Seidel:1990jh,Balakrishna:1997ej,Guzman:2004jw}).

We, however, focus on a particular subset of collapsing models.  Making use of the 
specific techniques developed by~\cite{Sanchis-Gual:2014nha} to evolve black-hole 
spacetimes in spherical symmetry using spherical-polar coordinates, we are able to 
follow the dynamics of the system for very long periods of time, well beyond black hole 
formation and in an entirely stable manner. Using these techniques we showed  
recently that quasi-bound states can form in the vicinity of a black hole born dynamically 
from the collapse of a neutron star surrounded by a scalar field~\cite{Sanchis-Gual:2015sxa}. 
Here we show that long-lived quasi-bound states can also form after the collapse of a 
self-interacting boson star. Similar results have also recently been obtained in 
spherical simulations of unstable Proca stars collapsing to black 
holes~\cite{sanchis2017numerical} as well as for axion stars~\cite{Helfer:2016ljl}.

This paper is organized as follows: Section~\ref{sec:formalism} briefly describes the 
mathematical formulation of the EKG system. Section~\ref{ID} discusses the construction 
of the initial data while Section~\ref{sec:NumImpl} gives a brief account of numerical 
aspects of the simulations. Our main findings and results are discussed in 
Section~\ref{sec:num_results}. Finally, Section~\ref{sec:conclusions} summarizes 
our conclusions. Greek indices run over spacetime indices while 
Latin indices run over spatial indices only. We use geometrized units, $c=G=1$.

\section{Basic equations}
\label{sec:formalism}

We investigate the dynamics of a self-interacting scalar field configuration around a black hole by solving numerically the coupled
EKG system 
\begin{equation}
 R_{\alpha\beta}-\frac{1}{2}g_{\alpha\beta}R=8\pi T_{\alpha\beta} \ ,
\label{eq:Einstein}
\end{equation}
with the scalar field matter content given by the stress energy tensor
\begin{eqnarray}
T_{\alpha\beta}&=&\frac{1}{2}(D_{\alpha}\Phi )^{*}(D_{\beta}\Phi )+\frac{1}{2}(D_{\alpha}\Phi)(D_{\beta}\Phi )^{*}\nonumber\\
&-&\frac{1}{2}g_{\alpha\beta}(D^{\sigma}\Phi)^{*}(D_{\sigma}\Phi)
-\frac{\mu^2}{2}g_{\alpha\beta}|\Phi \Phi^{*}| \ \nonumber \\
&-&\frac{1}{4}\,\lambda\,g_{\alpha\beta}|\Phi\Phi^{*}|^{2}.
\label{eq:tmunu}
\end{eqnarray}
We consider the following potential for the scalar field 
$V\bigl(\Phi^2\bigl)=\mu^2|\Phi|^2 +\frac{\lambda}{2}|\Phi|^4$, where 
$V_{\rm{int}}:=\frac{1}{4}\,\lambda\,|\Phi|^{4}$ is a quartic self-interaction 
potential with coupling $\lambda$. We also introduce the dimensionless quantity 
$\Lambda \equiv\lambda/4\pi \mu^2$. The scalar field obeys the Klein-Gordon 
equation
\begin{equation}
 \biggl(\Box -\frac{dV}{d|\Phi|^2}\biggl)\Phi=0 \ ,
\label{eq:KG}
\end{equation}
where the D'Alambertian operator is defined by $\Box:=
(1/\sqrt{-g})\partial_{\alpha}(\sqrt{-g}g^{\alpha\beta}\partial_{\beta})$. $\Phi$ is dimensionless and $\mu$ has dimensions of (length)$^{-1}$. 

In spherical symmetry, the spatial line element can be written as
\begin{equation}\label{isotropic}
 dl^2 = e^{4\chi } (a(t,r)dr^2+ r^2\,b(t,r)  d\Omega^2) \ ,
\end{equation}
where $d\Omega^2 = d\theta^2 + \sin^2\theta d\varphi^2$ is the solid angle element
and $a(t,r)$ and $b(t,r)$ are two non-vanishing metric functions. Moreover, $\chi$ is related to
the conformal factor $\psi$ as $\psi = e^\chi = (\gamma/\hat \gamma)^{1/12}$, with $\gamma$
and $\hat\gamma$ being the determinants of the physical and conformal 3-metrics, respectively. They 
are conformally related by $\gamma_{ij} = e^{4 \chi} \hat \gamma_{ij}$.

In this work we employ the Baumgarte-Shapiro-Shibata-Nakamura (BSSN) formalism
of Einstein's equations~\cite{Baumgarte98,Shibata95} where the evolved
fields are the conformally related 3-dimensional metric, the conformal exponent $\chi$, the trace of the extrinsic
curvature $K$, the independent component of the traceless part of the conformal extrinsic curvature, $A_{a}\equiv A^{r}_{\,\,r}, A_{b}\equiv A^{\theta}_{\,\,\theta}=A^{\varphi}_{\,\,\varphi}$,
and the radial component of the conformal connection functions. The interested reader is addressed to
\cite{Alcubierre:2010is,Montero:2012yr,sanchis2015quasistationary2} for further details. In particular, the explicit form of the evolution
equations for the gravitational field that we use in this work are given by Eqs.~(9)-(11) and (13)-(15) in Ref.~\cite{sanchis2015quasistationary2}.

As in our previous work~\cite{sanchis2016dynamical}, in order to solve the Klein-Gordon equation we use two first-order variables defined as
\begin{eqnarray}
\Pi &:=& n^{\alpha}\partial_{\alpha}\Phi=\frac{1}{\alpha}(\partial_{t}\Phi-\beta^{r}\partial_{r}\Phi) \ ,\\
\Psi&:=&\partial_{r}\Phi \ .
\end{eqnarray}
Therefore, from Eq.~(\ref{eq:KG}) we obtain the following system of first-order equations:
\begin{eqnarray}
\partial_{t}\Phi&=&\beta^{r}\partial_{r}\Phi+\alpha\Pi \ ,\\
\partial_{t}\Psi&=&\beta^{r}\partial_{r}\Psi+\Psi\partial_{r}\beta^{r}+\partial_{r}(\alpha\Pi) \ ,\\
\partial_{t}\Pi&=&\beta^{r}\partial_{r}\Pi+\frac{\alpha}{ae^{4\chi}}[\partial_{r}\Psi\nonumber\\
&+&\Psi\biggl(\frac{2}{r}-\frac{\partial_{r}a}{2a}+\frac{\partial{r}b}{b}+2\partial_{r}\chi\biggl)\biggl]\nonumber\\
&+&\frac{\Psi}{ae^{4\chi}}\partial_{r}\alpha+\alpha K\Pi - \alpha \bigl(\mu^{2}+\lambda \Phi^2\bigl)\Phi \ .
\label{eq:sist-KG}
\end{eqnarray}
The right-hand-sides of the gravitational field evolution equations contain matter source terms (see Eqs.~(9)-(11) and (13)-(15) in~\cite{sanchis2015quasistationary2}), denoted by $\mathcal{E}$, $S_{a}$, $S_{b}$ and $j_r$. These terms are components of the energy-momentum tensor, Eq.~\eqref{eq:tmunu}, or suitable projections thereof, and are given by
\begin{eqnarray}
\mathcal{E}&:=&n^{\alpha}n^{\beta}T_{\alpha\beta}=\frac{1}{2}\biggl(|\Pi|^{2}+\frac{|\Psi|^{2}}{ae^{4\chi}}\biggl) \nonumber\\
&&+\frac{1}{2}\mu^{2}|\Phi|^{2}+\frac{1}{4}\lambda\,|\Phi|^{4} \label{eq:rho}, \label{scalared}\\
j_{r}&:=&-\gamma^{\alpha}_{r}n^{\beta}T_{\alpha\beta}=-\frac{1}{2}\biggl(\Pi^{*}\Psi+\Psi^{*}\Pi\biggl)\ ,\\
S_{a}&:=&T^{r}_{r}=\frac{1}{2}\biggl(|\Pi|^{2}+\frac{|\Psi|^{2}}{ae^{4\chi}}\biggl) \nonumber\\
&&-\frac{1}{2}\mu^{2}|\Phi|^{2}-\frac{1}{4}\lambda\,|\Phi|^{4} \ ,\\
S_{b}&:=&T^{\theta}_{\theta}=\frac{1}{2}\biggl(|\Pi|^{2}-\frac{|\Psi|^{2}}{ae^{4\chi}}\biggl)  \nonumber\\
&&-\frac{1}{2}\mu^{2}|\Phi|^{2}-\frac{1}{4}\lambda\,|\Phi|^{4}\ .
\end{eqnarray}

The Hamiltonian and momentum constrains are given by the following two equations: 
\begin{eqnarray}
\mathcal{H}&\equiv& R-(A^{2}_{a}+2A_{b}^{2})+\frac{2}{3}K^{2}-16\pi \mathcal{E}=0,\label{hamiltonian}\\
\mathcal{M}_{r}&\equiv&\partial_{r}A_{a}-\frac{2}{3}\partial_{r}K+6A_{a}\partial_{r}\chi\nonumber\\
&+&(A_{a}-A_{b})\biggl(\frac{2}{r}+\frac{\partial_{r}b}{b}\biggl)-8\pi j_{r}=0.\label{momentum}
\end{eqnarray}
The latter two equations are only computed in our code to monitor the accuracy of the numerical evolutions.

\section{Initial data}
\label{ID}

Spherical boson stars are described by the radial function $\Phi(r,t) = \Phi_0(r) e^{i\omega t}$
where $\omega$ is the oscillation frequency of the field. Following~\cite{Colpi:1986ye} we obtain 
the initial data for a boson star in polar-areal coordinates, for which the line element is given by
\begin{equation}
  ds^2 = -\alpha^2(r')dt^2 + a^2(r')dr^2+r'^2d\Omega^2\ ,
\end{equation}
and $r'$ is the radial coordinate. The EKG system for a boson star reads:
\begin{eqnarray}
  \frac{\partial_{r'} a}{a} &=& \frac{1-a^2}{2r'}+2\pi r'\biggl[ \omega^2\Phi_0^2 \frac{a^2}{\alpha^2}+\Psi_0^2\nonumber\\
    &\quad&+a^2\Phi_0^2\left(\mu^2+\frac{1}{2}\lambda\Phi_0^2\right)\biggl]\ ,\label{eq:boson_aaa}\\
  \frac{\partial_{r'} \alpha}{\alpha} &=& \frac{a^2 - 1}{r'}\nonumber\\
  &\quad&+\frac{\partial_{r'} a}{a}-4\pi  r' a^2\Phi_0^2\left(\mu^2+\frac{1}{2}\lambda\Phi_0^2\right)\ ,\label{eq:boson_phi}\\
  \partial_{r'} \Phi_0 &=& \Psi_0\ ,\label{eq:boson_phi}\\
  \partial_{r'} \Psi_0 &=& -\Psi_0\left(\frac{2}{r'} + \frac{\partial_{r'} \alpha}{\alpha}- \frac{\partial_{r'} a}{a}\right) \nonumber\\
  &\quad&-\omega^2\Phi_0^2\frac{a^2}{\alpha^2} + a^2(\mu^2 + \lambda\Phi_0^2)\Phi_0\ .\label{eq:boson_psi}
\end{eqnarray}

\begin{figure}
  \begin{minipage}{1\linewidth}
    \includegraphics[width=1.02\textwidth]{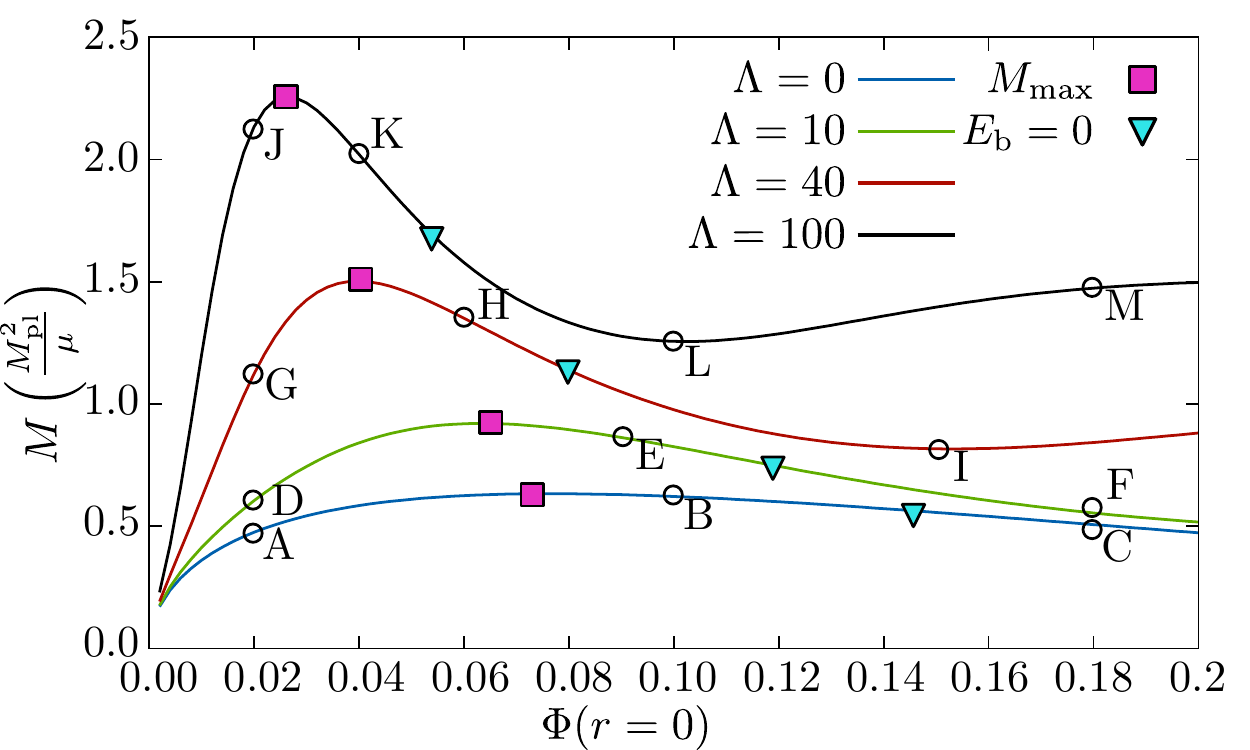} 
    \caption{Boson star mass  as a function of the central value of the scalar 
    field for a few sequences of equilibrium models with $\Lambda= \lbrace0, 10, 
40, 100\rbrace$. For each value of $\Lambda$ the purple squares indicate the 
maximum mass of the sequence, separating the stable and unstable branches,  and 
the inverted cyan triangles indicate the point at which the binding energy is 
zero. The empty circles (and letters) indicate the specific models we evolve in 
this work.}
    \label{fig:masses}
  \end{minipage}
\end{figure}

\begin{table*}
\caption{Initial parameters for the bosons stars with $\mu=1$. $\Lambda$ is  
the self-interaction coupling constant, $\Phi(r=0)$ is the central value of the scalar field, $R$ is 
the radius, $M_{\rm{MS}}$ is the Misner-Sharp mass, $M_{\rm{BS}}$ is the scalar-field total 
mass, $N$ is the bosonic number, $E_{b}$ is the binding energy 
and $\omega$ is the frequency. Models A, D, G, and J are stable; models B, E, H, and K 
can collapse to a black hole or migrate depending on the perturbation; and  
models C, F, I, L, and M disperse away.}
\label{tab:dades}
\begin{ruledtabular}
      \begin{tabular}{ccccccccc}
        Model &$\Lambda$ &$\Phi(r=0)$ &$R$ &$M_{\rm{MS}}$ &$M_{\rm{BS}}$ &$N$ 
&$E_{\text{b}} = M_{\rm{MS}}-N\mu$ &$\omega$\\
        \hline
      A &0   &$0.02$  &68.79 &0.47514 &0.47517 &0.48197 &-0.00683 &0.95394\\ 
      B &0   &$0.10$  &32.00 &0.62180 &0.62029 &0.63736 &-0.01556 &0.82269\\ 
      C &0   &$0.18$  &28.18 &0.50671 &0.49702 &0.47201 & 0.03470 &0.76883\\
      D &10  &$0.02$  &64.55 &0.60425 &0.60429 &0.61585 &-0.01160 &0.94580\\ 
      E &10  &$0.09$  &30.76 &0.86314 &0.86103 &0.89092 &-0.02778 &0.79838\\ 
      F &10  &$0.18$  &29.61 &0.55419 &0.53416 &0.46712 & 0.08707 &0.81866\\ 
      G &40  &$0.02$  &57.72 &1.12319 &1.12327 &1.16316 &-0.03997 &0.91481\\ 
      H &40  &$0.06$  &30.39 &1.35099 &1.35036 &1.40322 &-0.05223 &0.79616\\ 
      I &40  &$0.15$  &33.13 &0.81671 &0.79145 &0.70741 & 0.10930 &0.86111\\ 
      J &100 &$0.02$  &45.77 &2.13376 &2.13398 &2.25789 &-0.12413 &0.86267\\ 
      K &100 &$0.04$  &29.53 &2.02134 &2.02138 &2.10953 &-0.08819 &0.79656\\ 
      L &100 &$0.10$  &30.50 &1.26039 &1.23670 &1.14872 & 0.11167 &0.85837 \\
      M &100 &$0.18$  &30.70 &1.47383 &1.45448 &1.38584 & 0.08799 &0.85576 \\ 
    \end{tabular}
\end{ruledtabular}
\end{table*}

By solving these equations we obtain the spacetime metric potentials, 
$g_{tt}=-\alpha^2, g_{rr}=a^2$, and the  radial distribution of the scalar 
field, $\Phi_0$. The mass of a boson star is computed using  the definition of 
the Misner-Sharp mass function,
\begin{eqnarray}
  M_{\rm{MS}} = \frac{r'_{\text{max}}}{2}\left( 1 - 
\frac{1}{a^2(r'_{\text{max}})}\right)\ ,\label{eq:massa_adm_sch}
\end{eqnarray}
where $r'_{\text{max}}$ is the radial coordinate at the outer boundary of our computational grid.
The total mass of a boson star can also be computed using the Komar integral~\cite{herdeiro2016kerr},
\begin{equation}\label{eq:massa_kg}
  M_{\rm{BS}} = \int_{\Sigma} 
\left(2T_t^t-T_\alpha^\alpha\right)\alpha\sqrt{\gamma} \,dr\,d\theta\, 
d\varphi\ ,
\end{equation}
where $\Sigma$ is a spacelike slice extending from a horizon, in case one exists, 
up to spatial infinity. To study the stability properties of the constructed 
equilibrium models we need to compute the Noether charge associated with the total
bosonic number $N$, which is defined as
\begin{equation}\label{eq:nombre_bosons}
  N = \int g^{0\nu}j_{\nu}\alpha\sqrt{\gamma} \,dr\,d\theta\, 
d\varphi\ ,
\end{equation}
where $j_{\nu} = \frac{i}{2}(\Phi^*\partial_\nu\Phi - \Phi\partial_\nu\Phi^*)$ 
is the conserved current associated with the transformation of the U(1) group. 
Finally, the sign of the binding energy
\begin{equation}\label{binding}
E_{\text{b}} = M_{\rm MS}-N\mu,
\end{equation}
will determine the outcome of unstable models.

Representative sequences of equilibrium models of boson stars are plotted in 
Figure~\ref{fig:masses}. This figure shows four different mass profiles as a 
function of the central scalar field value for four different values of the 
self-interaction coupling constant, namely $\Lambda =\lbrace0,10,40,100\rbrace$. 
For any given $\Lambda$ two important points are explicitly indicated in each 
curve, the maximum mass, marked with a purple square, and the point at which 
$E_{\text{b}} = 0$, marked with a cyan inverted triangle. For each sequence, the 
location of the maximum mass indicates the critical point separating the stable 
and the unstable branches. Boson stars situated at the left of the point of 
maximum mass are stable, while those on the right are unstable. The maximum mass 
increases monotically with $\Lambda^{1/2}$~\cite{liebling2012dynamical}. For 
sufficiently large values of $\Lambda$ the self-interaction term allows for 
significantly larger masses than for non-self-interacting (mini-)boson stars.
 
We study the stability of these equilibrium models through numerical time 
evolutions. These are triggered by adding suitable small-amplitude perturbations 
to the initial data profiles. We consider two types of perturbations, either 
those associated with the intrinsic truncation error of the finite-difference 
representation of the PDEs we solve or those associated with a functional 
modification of the actual radial distributions. While the evolutions of the 
boson stars may seem a priori easily predictable, telling from their location 
with respect to the maximum in the $M_{\rm BS}$ vs. $\Phi(r=0)$ diagram, 
there are other aspects to consider which may affect the actual evolutions. In 
fact, depending on the sign of the binding energy, the point at which 
$E_{\text{b}} = 0$, and on the perturbation, the stars will undergo different 
fates. On the one hand, as we show below, an unstable boson star with positive 
binding energy which is perturbed only with the discretization numerical error, 
migrates to the model with the same mass in the corresponding stable branch. 
However, if it is perturbed by slightly increasing its mass, it can collapse 
gravitationally and form a black hole. On the other hand, a boson star with an 
excess energy, i.e.~placed at the right of the zero binding energy point, is no 
longer bounded and it will disperse away with time. 
 
The specific boson-star models that we generate and evolve numerically are indicated by the empty circles in Fig.~\ref{fig:masses}. Quantitative details of the main model parameters are reported in Table~\ref{tab:dades}. The boson star initial configurations are built in polar-areal coordinates but the time evolutions are performed in our code using isotropic coordinates, Eq.~(\ref{isotropic}). To solve this problem, we have to take two steps, see~\cite{liebling2012dynamical,lai2004numerical}. First, we perform a change of coordinates from polar-areal to isotropic coordinates with
\begin{eqnarray}
  r_{\text{max}} &=& \left[ \left(\frac{1+\sqrt{a(r'_{\text{max}})}}{2}\right)^2\frac{r'_{\text{max}}}{a(r'_{\text{max}})}\right] \ , \label{eq:rmax_inte}\\
  \frac{dr}{dr'} &=& a\frac{r}{r'}\ , \label{eq:r_iso_sch}
\end{eqnarray}
where Eq.~(\ref{eq:rmax_inte}) is used as the initial value to integrate Eq.~(\ref{eq:r_iso_sch}) backwards. Then, we obtain the conformal factor using
\begin{equation}
\psi = \sqrt{\frac{r'}{r}}\ .
\end{equation}
With this procedure our initial solution is described in isotropic coordinates and we can write the initial values of the other scalar field quantities as:
\begin{eqnarray}
  \Phi(r, t = 0) &=& \Phi_0\ ,\label{eq:ce_phi}\\
  \Psi(r, t = 0) &=& \Psi_0\ , \\
  \Pi(r, t = 0)  &=& i\frac{\omega}{\alpha}\Phi_0\ .\label{eq:ce_pi}
\end{eqnarray}
Finally, we interpolate the solution in an isotropic grid employing a 
cubic-spline interpolation~\cite{press1992numerical} that guarantees the 
continuity of the second derivative to minimize high-frequency noise associated 
with the interpolation. 

\section{Numerical framework}
\label{sec:NumImpl}

As in our previous papers, the BSSN evolution equations for the geometry and 
the evolution equations for the scalar field are solved numerically using a 
second-order PIRK scheme~\cite{Isabel:2012arx,Casas:2014}. This scheme can 
handle in a satisfactory way the singular terms that
appear in the evolution equations due to our choice of curvilinear coordinates. 
Explicit details about our numerical implementation have been reported 
e.g.~in~\cite{sanchis2016dynamical}. We also note 
that the convergence properties of our numerical code have been extensively 
tested before in various physical systems, including the EKG equations with 
self-interaction, see 
e.g.~\cite{sanchis2016dynamical,sanchis2015quasistationary, 
sanchis2015quasistationary2,sanchis2016explosion}.

In the simulations reported in this work we consider two computational grids, 
namely a grid to obtain the equilibrium models of boson stars in polar-areal 
coordinates, and another one to evolve those models in isotropic coordinates. 
Our polar-areal grid is an equidistant grid with spatial resolution $\Delta r' = 
0.001$ spanning the interval $r' \in [0.0,150.0]$. On the other hand, our 
isotropic grid is composed of two patches, a geometrical progression in the 
interior part up to a given radius and a hyperbolic cosine outside. Using the 
inner grid alone would require too many grid points to place the outer boundary 
sufficiently far from the origin (and hence prevent the effects of possible 
spurious reflections), while using only the hyperbolic cosine patch would 
produce very small grid spacings in the inner region of the domain, leading to 
prohibitively small timesteps due to the Courant-Friedrichs-Lewy (CFL) 
condition. Details about the computational grid can be found in 
\cite{sanchis2015quasistationary}. In our work the minimum resolution $\Delta r$ 
we choose for the isotropic logarithmic grid is $\Delta r=0.025$. With this 
choice the inner boundary is then set to $r_{\text{min}} = 0.0125$ and the outer 
boundary is placed at $r_{\text{max}} = 6000$ at the nearest (in some models it 
is placed even further away, at $r_{\text{max}} = 10000$). The time step is 
given by $\Delta t = 0.3 \Delta r$ in order to obtain long-term stable 
simulations. 

\begin{figure}
\begin{minipage}{1\linewidth}
\includegraphics[width=1.0\textwidth, height=0.22\textheight]{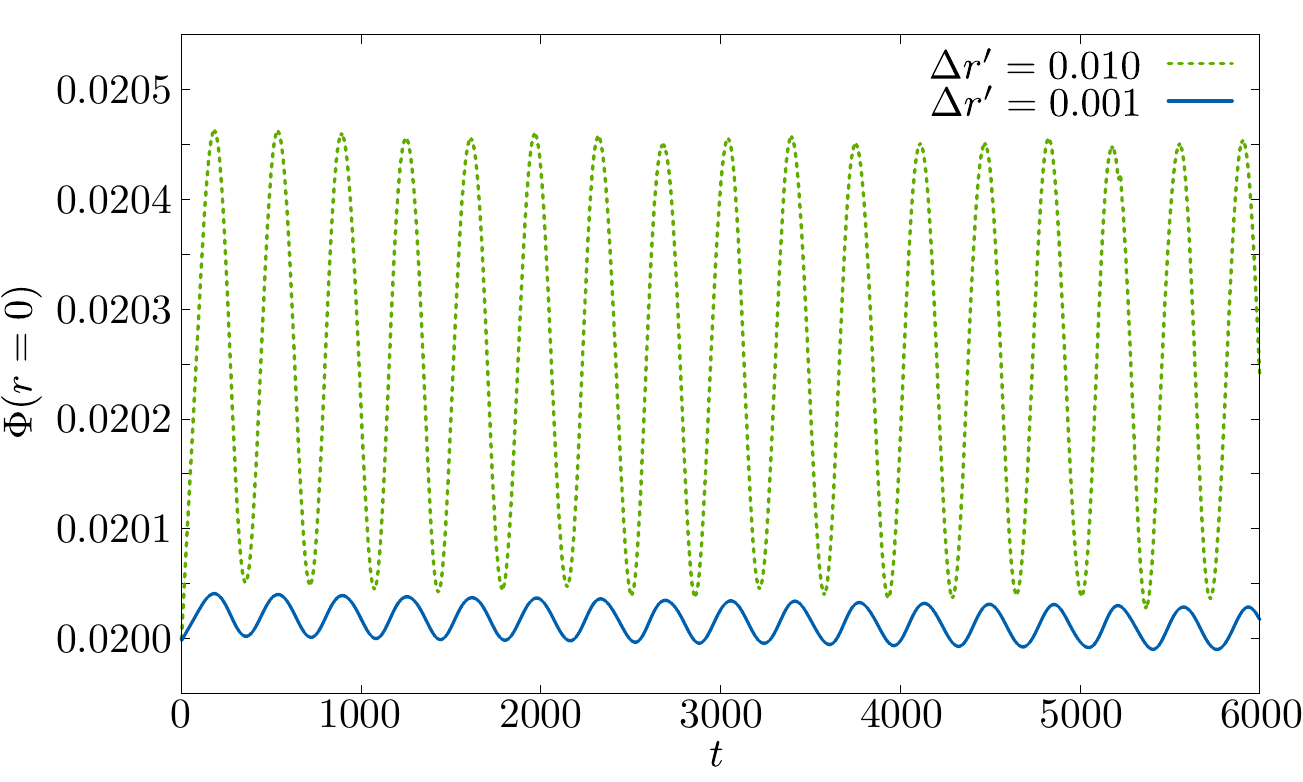} 
\caption{Time evolution of the central value of the scalar field for 
boson star model A ($\Phi(r=0, t=0) = 0.02$ and $\Lambda = 0$) for two different 
resolutions of the initial data grid ($\Delta r'$).}
\label{fig:Acomp}
\end{minipage}
\end{figure}

\section{Results} \label{sec:num_results}

\subsection{Stable models}

\begin{figure}
\begin{minipage}{1\linewidth}
\includegraphics[width=1.0\textwidth, height=0.22\textheight]{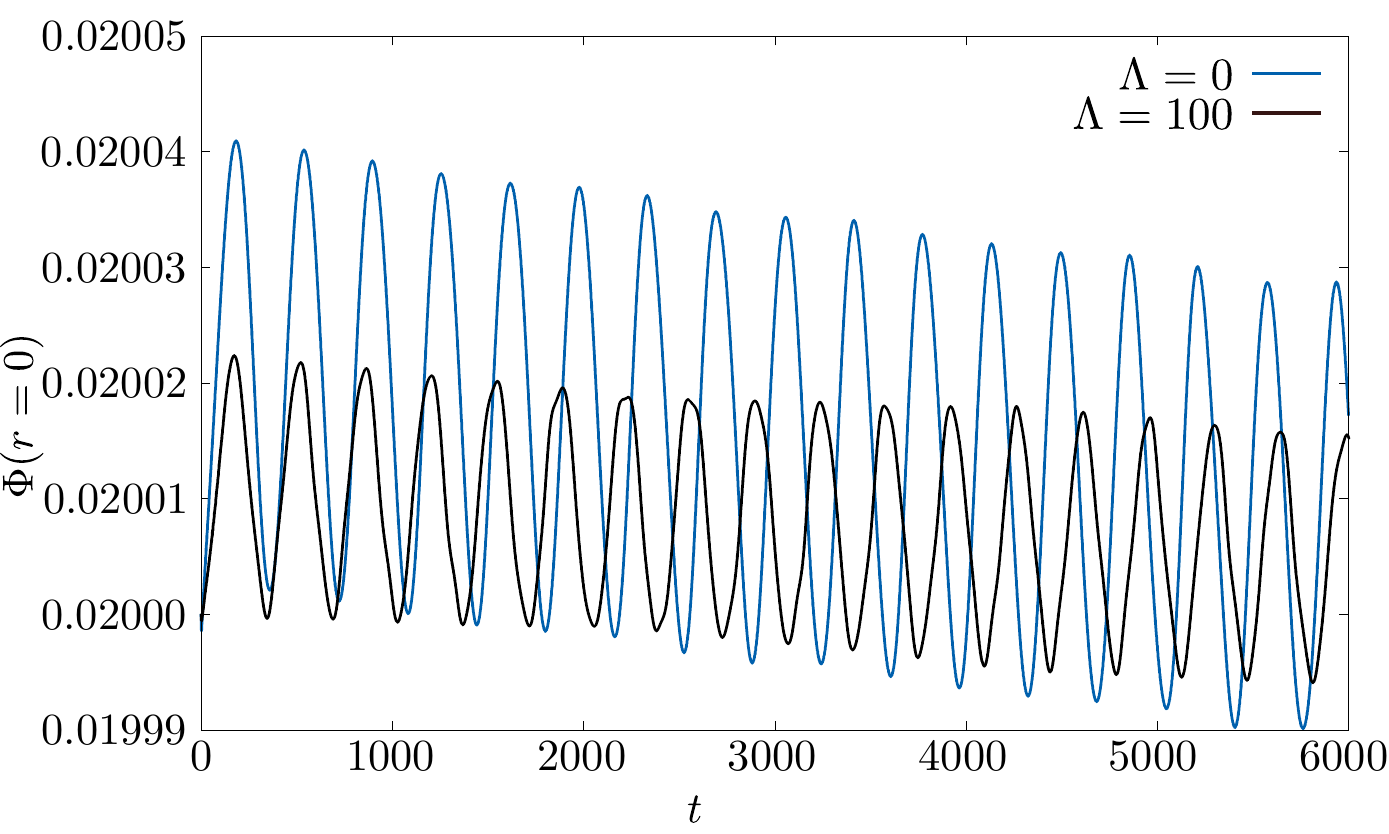} 
\caption{Time evolution of the central value of the scalar field for a 
boson star with $\Phi(r=0, t= 0) = 0.02$ and two different values of the self-interaction 
coupling constant $\Lambda = \{0, 100\}$, corresponding to models A and J, respectively. 
The evolutions were done with the grid resolution of the initial data $\Delta r'=0.001$.}
\label{fig:estables}
\end{minipage}
\end{figure}

Models A, D, G, and J in Fig.~\ref{fig:masses} are all stable models. 
Therefore, the time evolution of the physical quantities that characterize them, 
as e.g.~the central value of the scalar field, should remain constant. However, 
due to the grid discretization error, any of those quantities will instead 
oscillate around the equilibrium value. This is shown in Fig.~\ref{fig:Acomp} 
where we plot the central value of the scalar field for model A for two 
different resolutions of the polar-areal grid employed to build the initial 
data. In this figure we can also observe how when the resolution of the initial 
data is reduced from $\Delta r'=0.01$ to $\Delta r'=0.001$ the amplitude of the 
oscillation is significantly reduced. All of our stable initial models are 
indeed seen to oscillate around the central equilibrium values. As an example of 
their stability we plot in Fig.~\ref{fig:estables} the time evolution of the 
central scalar field for models A ($\Lambda =0$) and J ($\Lambda =100$). 
Note that the (purely numerical) secular drift of the initial central value of the scalar field of 
model A  apparent in Fig.~\ref{fig:estables} and hardly visible in Fig.~\ref{fig:Acomp}
(compare the two blue curves, both corresponding to the model with $\Lambda=0$) is 
simply a consequence of the change of scale 
in the vertical axes of both figures. By Fourier-transforming the time evolution 
of the central value of the scalar field we obtain the corresponding frequency of 
oscillation $\omega$ of the models. Those values are reported in the last column of 
Table~\ref{tab:dades}. The stable models oscillate with a single fundamental 
frequency whose value decreases with increasing $\Lambda$.

\subsection{Unstable models}

\begin{figure}
\begin{minipage}{1\linewidth}
\includegraphics[width=1.0\textwidth, height=0.22\textheight]{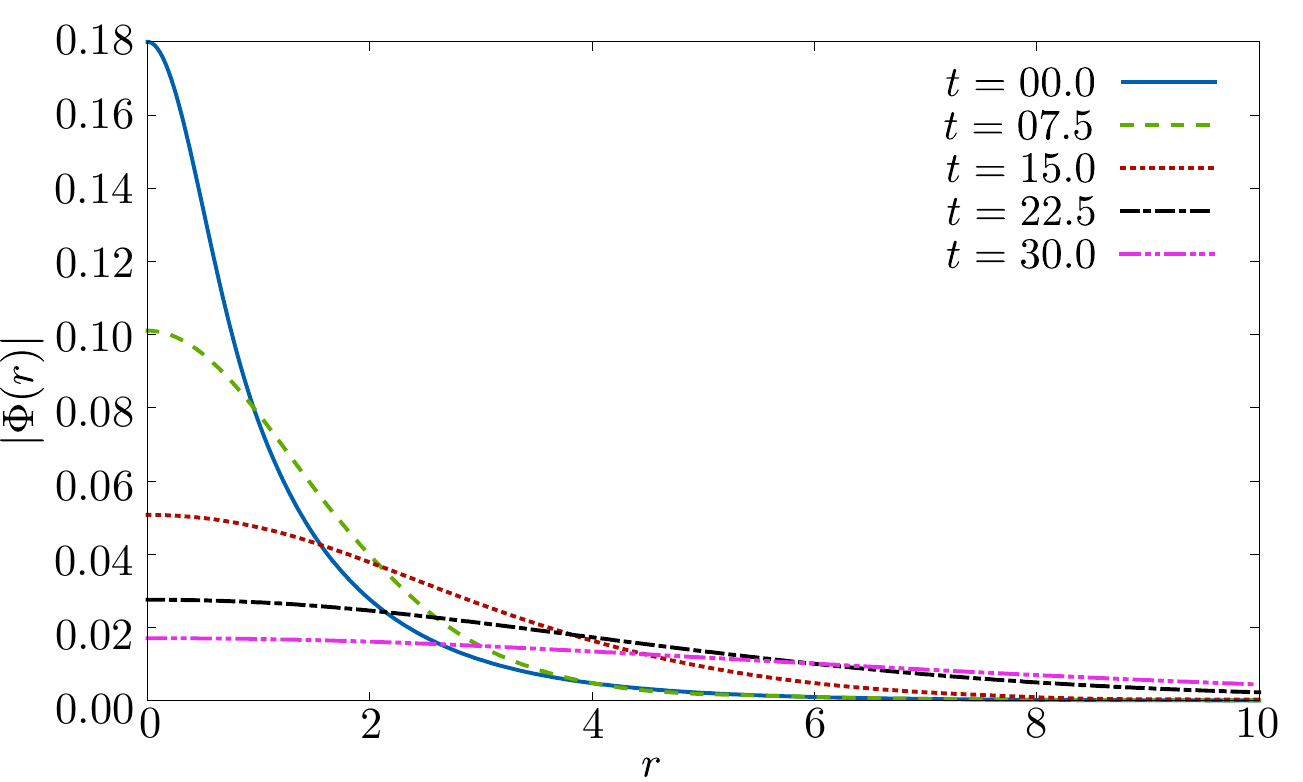} 
\caption{Radial profile of the scalar field amplitude for different times, 
$t = \{0.0, 7.5, 15.0, 22.5, 30.0\}$, with $\Phi(r=0,t=0) = 0.18$ and $\Lambda = 
0.0$ corresponding with the C case.}
\label{fig:Cphi_evol}
\end{minipage}
\end{figure}

Another possible outcome of the evolution of our initial data is the total 
dispersion of the boson star or its gravitational collapse. Let us start 
considering the first possibility. Such unstable situation will happen when the 
binding energy is positive since due to the energy excess the 
star will no longer remain bounded. The subset of initial models that can follow 
this trend are boson stars C, F, I, L, and M in Fig.~\ref{fig:masses}. As an 
example we plot in Fig.~\ref{fig:Cphi_evol} the radial profiles of the scalar 
field at selected times of the evolution corresponding to model C (indicated in 
the legend). In this case the central value of the scalar field rapidly 
decreases with time, the boson star suffers a drastic radial expansion and 
disperses away. All other unstable models (F, I, L and M) with a positive 
binding energy display the same fate.

\begin{figure}
\begin{minipage}{1\linewidth}
\includegraphics[width=1.0\textwidth, height=0.22\textheight]{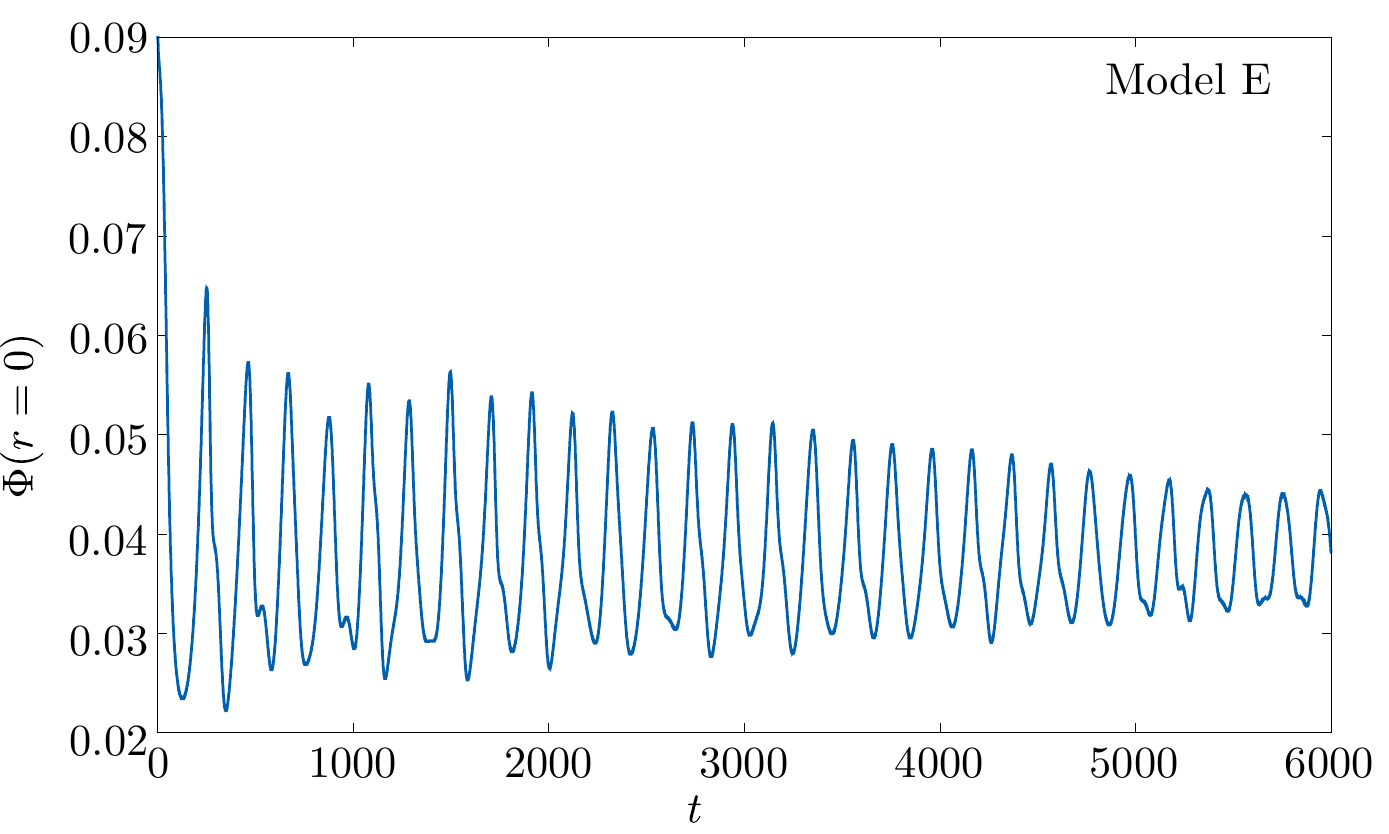}
\includegraphics[width=1.0\textwidth, height=0.22\textheight]{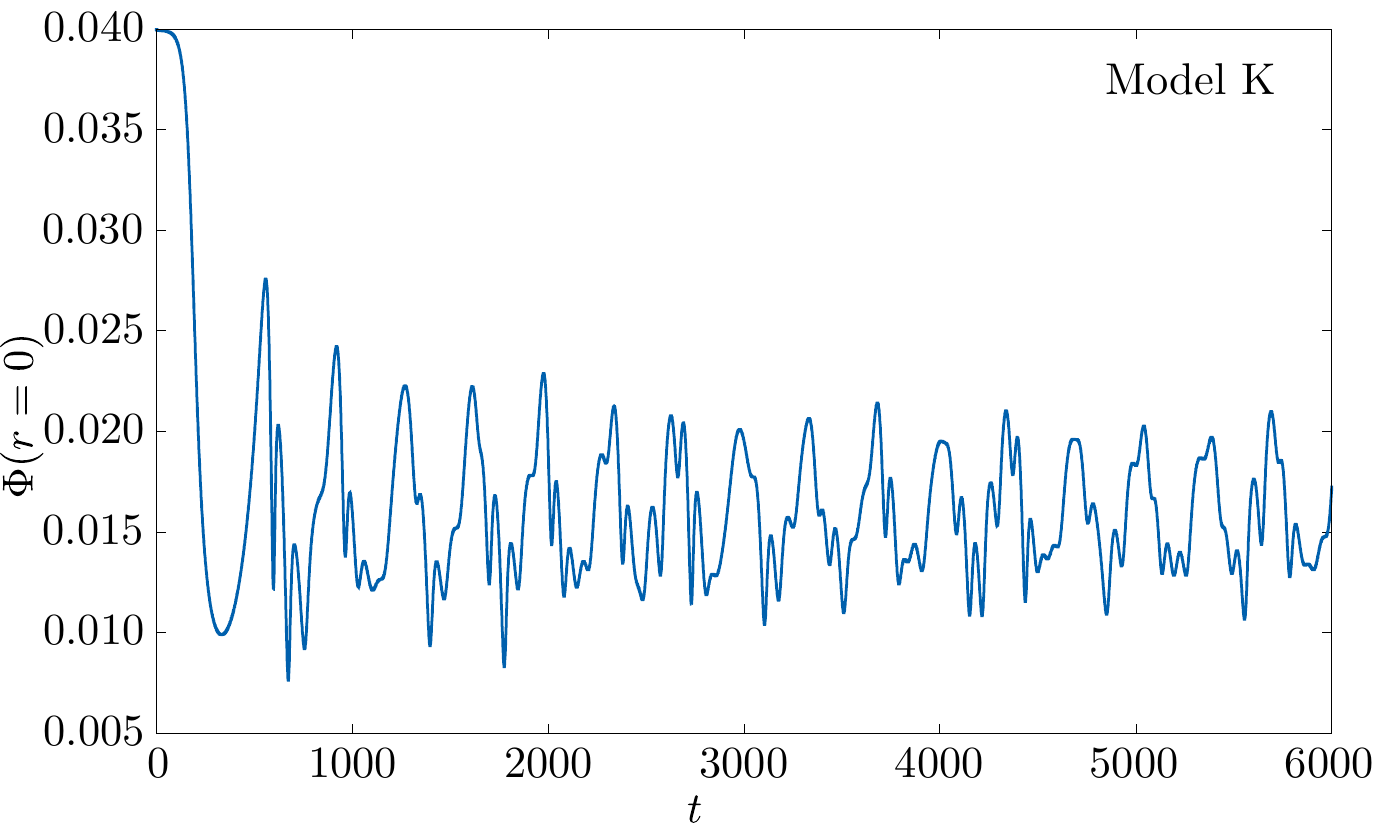}  
\caption{Time evolution of the central value of the scalar field for 
the non-perturbed unstable boson star models E ($\Lambda = 
10$; top panel) and K ($\Lambda = 100$; bottom panel). The stars migrate 
to the stable branch and oscillate around the value corresponding to models 
with the same mass.}
\label{fig:Ephi}
\end{minipage}
\end{figure}

Let us now consider the evolution of initial models that are located in the 
unstable branch, i.e.~between the critical point (maximum mass of the 
configuration) and the $E{_\text{b}}= 0$ point. These are models B, E, H, and K 
in Fig.~\ref{fig:masses}. Previous numerical work~\cite{guzman2009three} has 
shown that the fate of these models is to collapse gravitationally to form a 
black hole. However, we find that these models can also {\it migrate} to the 
stable branch of equilibrium configurations,  depending on the perturbation (see 
also earlier work by~\cite{Seidel:1990jh}). If the only perturbation of the 
initial data is the one due to the discretization error, the outcome is a 
migration to the stable branch. However, if we include a slightly larger 
perturbation in the initial data, the models collapse to form black holes. The 
first type of evolution, while mathematically plausible but unlikely on physical 
grounds, has been previously observed in the case of neutron stars 
(see~\cite{Font:2002} for details and arguments against this evolution), boson 
stars~\cite{Seidel:1990jh} and, recently, also in the case of unstable Proca 
stars~\cite{sanchis2017numerical}. A migrating boson star will result in a 
different boson star. It will have the same mass but it will be located in the 
stable branch of the equilibrium configurations (and hence the central scalar 
field will have a smaller value). As an example, Fig.~\ref{fig:Ephi} shows 
the migration of boson star models E and K. For model E the star moves from a 
central value of the scalar field $\Phi(r=0, t=0) = 0.09$ towards a final value of 
$\Phi(r=0) \sim 0.04$. This is precisely the value for which we obtain a stable 
boson star with the same mass (cf.~Fig.~\ref{fig:masses}). In the bottom panel 
of Fig.~\ref{fig:Ephi} one can also observe that, besides the overall migration, 
the evolution of model K excites more frequencies of oscillation than that of 
model E. This behaviour is consistent with the fact that the nonlinear term in 
the potential induces nonlinear couplings among the frequencies, which are 
more apparent the larger the value of the self-interaction coupling constant. 

Let us now consider the evolution of truly perturbed unstable models. 
To perturb these models we add an extra $2\%$ value to the initial scalar field 
by multiplying $\Phi$ by 1.02 after solving equations (\ref{eq:boson_aaa})-(\ref{eq:boson_psi}).
We have checked by computing the binding energy that the perturbation does not change 
the sign. We then compute the auxiliary variables given by Eqs.~(\ref{eq:ce_phi})-(\ref{eq:ce_pi}) 
using the perturbed scalar field. For simplicity, after adding the perturbation we do not 
recompute the spacetime variables $a$ and $\alpha$. This produces a slight violation 
of the constraints and leads to a small difference between the masses computed with Eq.~(\ref{eq:massa_kg}) and  Eq.~(\ref{eq:massa_adm_sch}) in polar-areal coordinates. 
However, since the perturbation is fairly small (yet still larger than that associated with the 
discretization error) it does not substantially alter our original solution.

\begin{figure}[t]
  \begin{center}
    \includegraphics[width=0.48\textwidth]{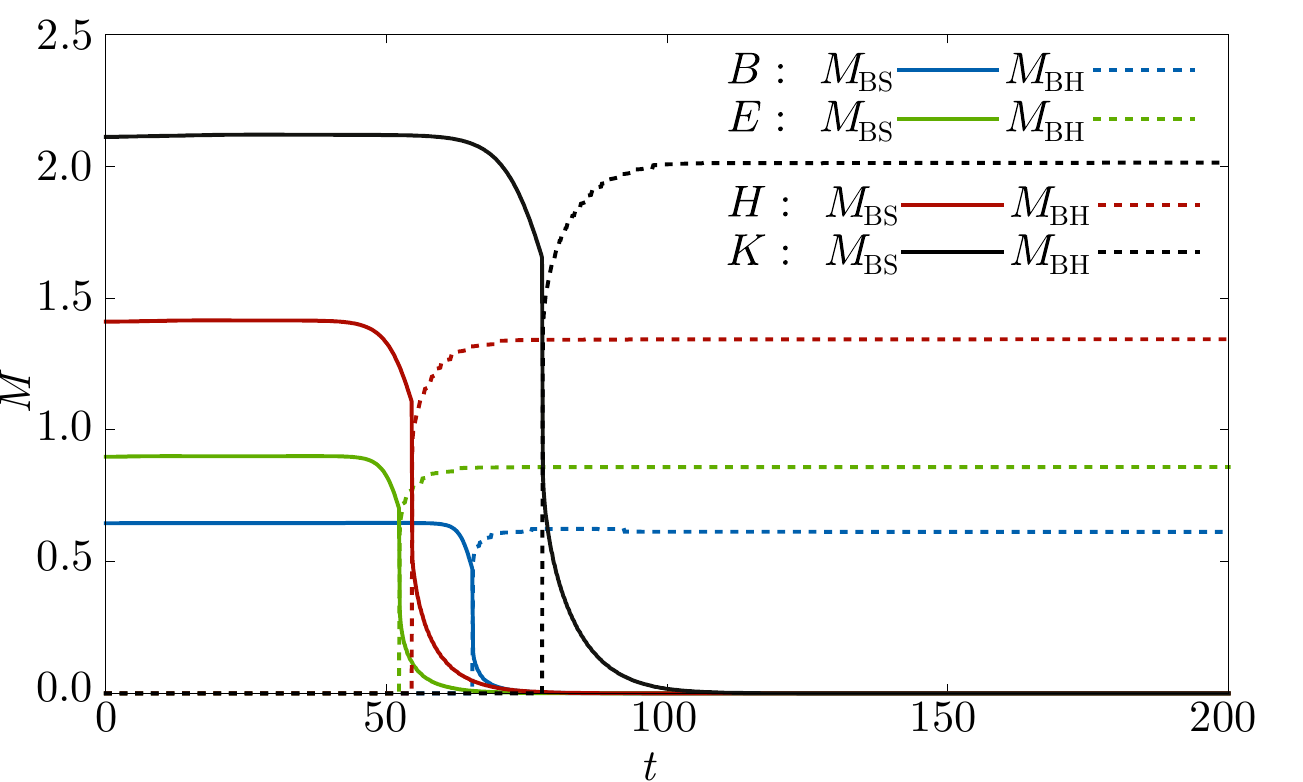}
    \caption{Time evolution of the scalar field mass/energy (continuous lines) 
    and the BH mass (dashed lines) for the four different unstable models B, E, 
H and K after the introduction of a 2\% perturbation in the initial data.}
    \label{fig:masses_forat}
  \end{center}
\end{figure}

To diagnose the appearance of a black hole in the evolution we compute the 
mass of the BH through the apparent horizon (AH) area $\mathcal{A}$, using 
$M_{\text{BH}}=\sqrt{\mathcal{A}/16\pi}$. The time 
evolution of both the scalar 
field energy (mass) and the BH mass for all unstable models is shown in 
Fig.~{\ref{fig:masses_forat}. The mass of the boson star is computed using the 
Komar integral, Eq.~(\ref{eq:massa_kg}). Contrary to the migrating case, adding 
a 2\% perturbation on the initial data triggers the collapse of the solutions 
and at some point in the evolution an AH forms. This time is indicated in 
Fig.~\ref{fig:masses_forat} by the sudden change that is observed in the 
evolution of the energy of the scalar field, which is associated with the sudden 
increase of the black hole mass from a zero value.

Fig.~\ref{fig:masses_forat} shows that, as expected, there is a small difference 
between the boson star masses computed with Eq.~(\ref{eq:massa_kg}) 
and the black hole masses computed through the apparent horizon. This is 
because some part of the scalar field is released after the collapse. 
For all models, the black hole mass is consistently, and slightly, smaller (see 
Table~\ref{tab:dades2}). This means that during the collapse 
to a black hole, a remnant of the initial scalar field is not swallowed by the 
hole but instead lingers around in the form of a spherical shell or cloud. 
Figs.~\ref{fig:B_p} and~\ref{fig:H_p} show the time evolution of the amplitude 
of the central value of the scalar potential for the four unstable models. This 
time series is extracted at an observation point with a fixed radius 
$r_{\rm{obs}}= 10$.  The scalar field does not disappear after the formation of 
the AH (which takes place for all models (well) before $t=100$; 
cf.~Fig.~\ref{fig:masses_forat}), forming instead long-lived quasi-bound states. 
For  all  of  the  models  the   field  is  seen  to  be clearly oscillating, as 
it is best visualized in the insets of the two figures. To identify the 
frequencies at which the  field oscillates we perform a Fourier transform of the 
time series and obtain the power spectrum. This power spectrum shows a set of 
distinct frequencies, as indicated in Table~\ref{tab:dades2}. Moreover, models 
B, E, and H, show a distinctive beating pattern due to the presence of overtones 
of the fundamental frequency. 

\begin{figure}[t!]
  \begin{center}
    \includegraphics[width=0.47\textwidth]{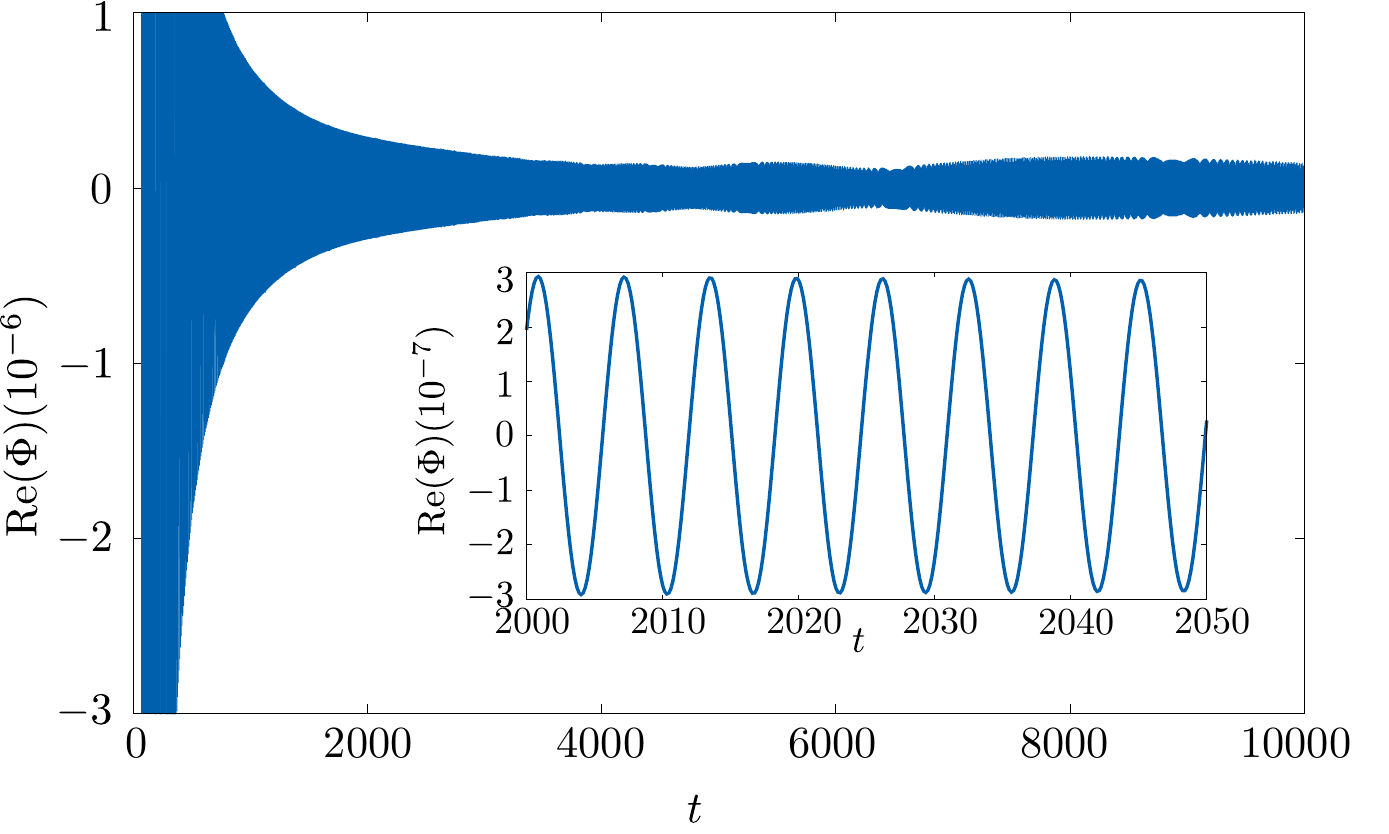}
       \includegraphics[width=0.47\textwidth]{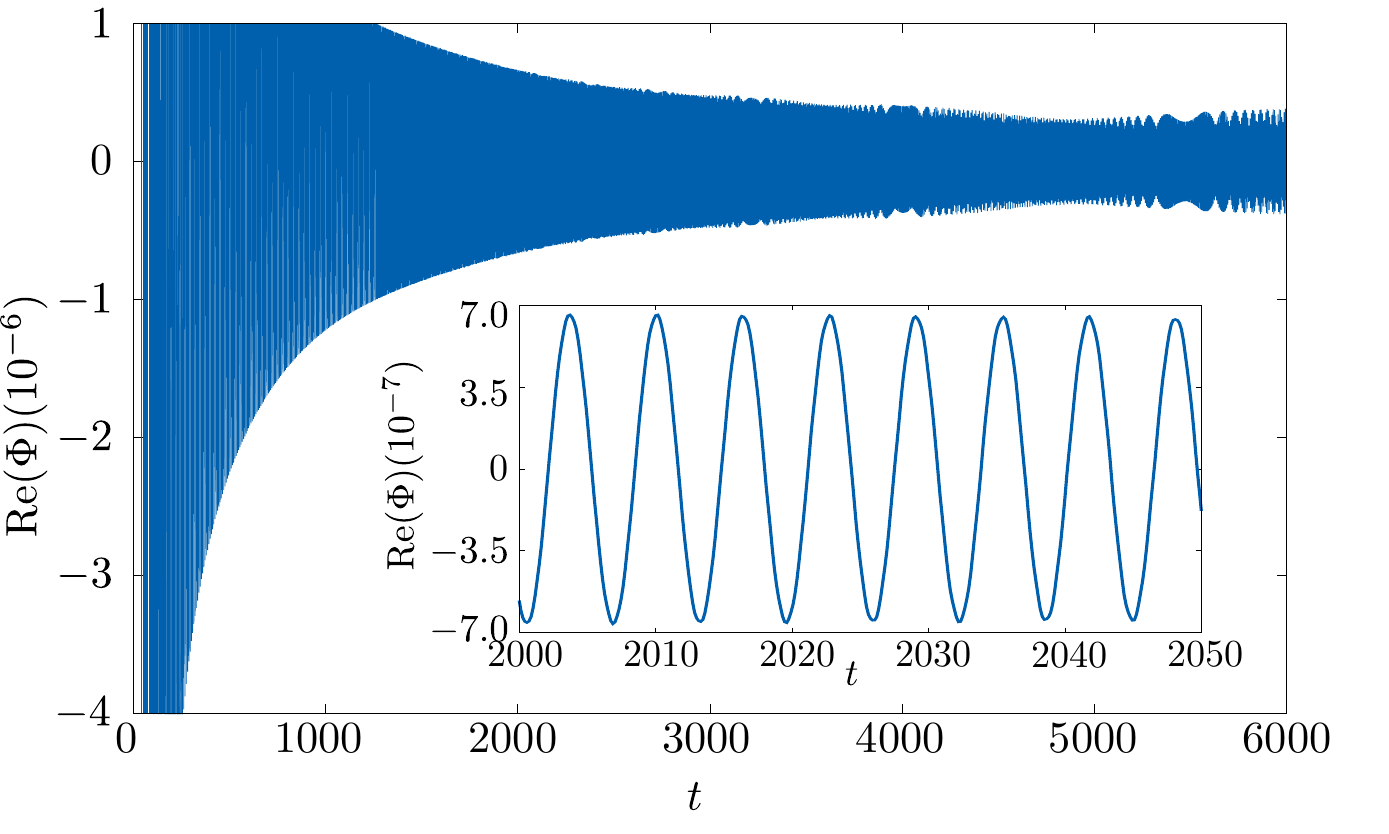}
    \caption{Time evolution of the real part of the central value of 
    the scalar field for the perturbed boson star models B (top panel) and E 
(bottom panel). The insets show a magnified view of $t\in[2000,2050]$\ in the 
evolution to highlight the oscillatory behaviour of the scalar field that 
lingers outside of the black hole.}
    \label{fig:B_p}
  \end{center}
\end{figure}

\begin{figure}[t!]
  \begin{center}
    \includegraphics[width=0.47\textwidth]{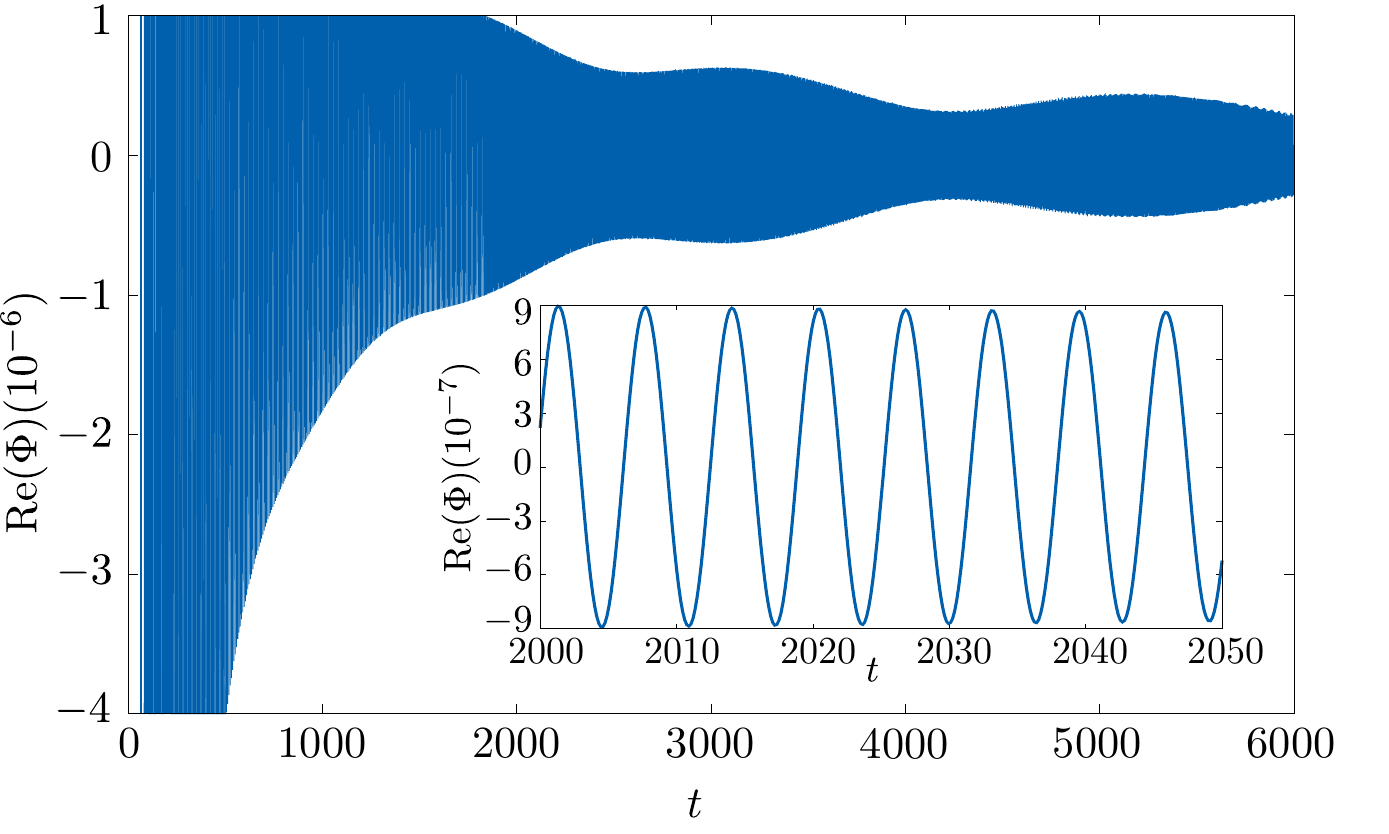}
      \includegraphics[width=0.47\textwidth]{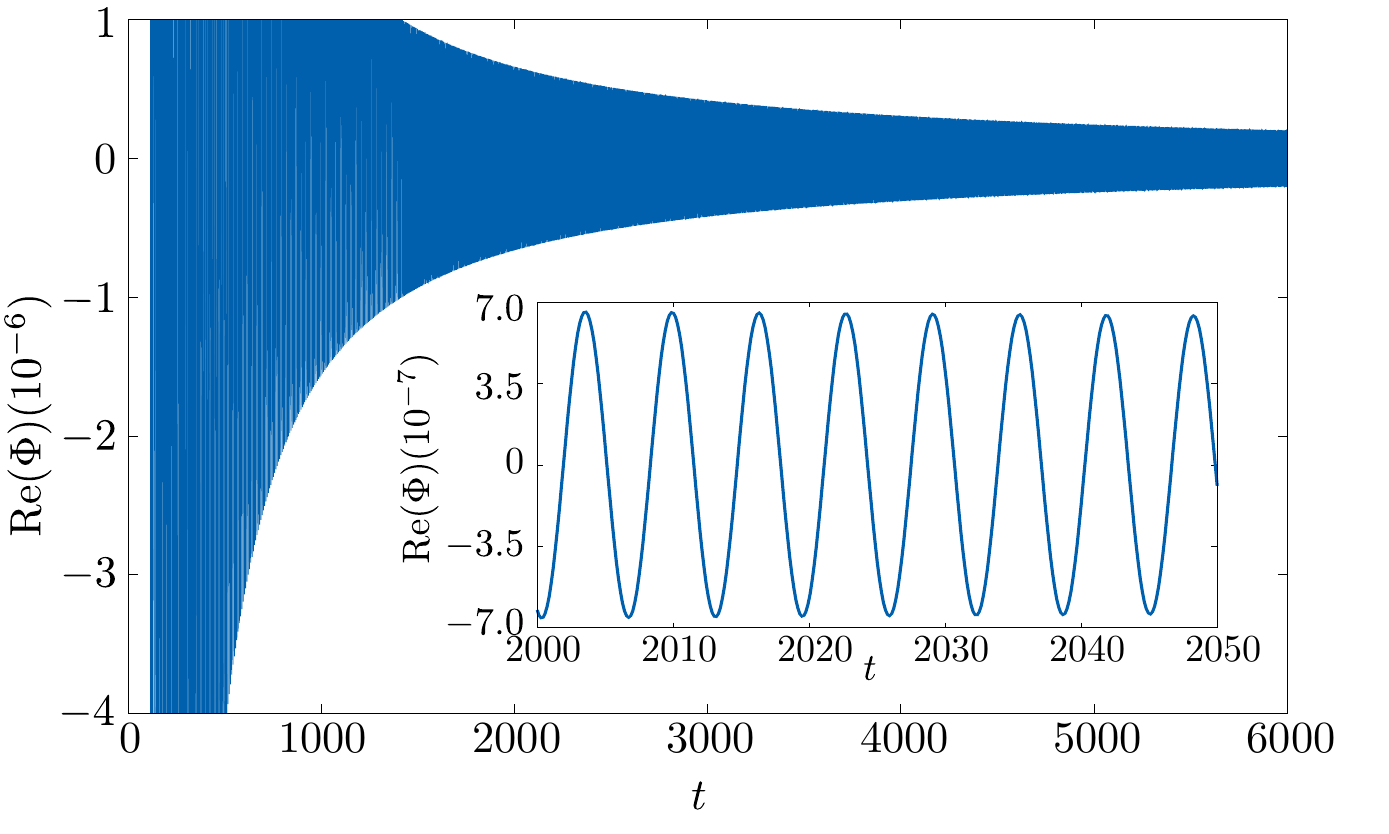}
    \caption{Same as Fig.~\ref{fig:B_p} but for models H (top panel) and K (bottom panel).}
    \label{fig:H_p}
  \end{center}
\end{figure}

\begin{table}
\caption{Masses of unperturbed boson stars, $M_{\rm{BS}}$, masses of the 
black holes formed following boson star collapse, $M_{\rm{BH}}$, and frequencies of the 
scalar-field quasi-bound states, either computed after black hole formation, $\omega_{\text{qb}}^{(1)}$,
or computed  from the scattering of a Gaussian pulse, $\omega_{\text{qb}}^{(2)}$.}
\label{tab:dades2}
\begin{ruledtabular}
      \begin{tabular}{ccccccc}
        Model &$\Lambda$ &$\Phi(r=0)$ &$M_{\rm{BS}}$ & $M_{\rm{BH}}$ &$\omega_{\text{qb}}^{(1)}$&$\omega_{\text{qb}}^{(2)}$\\
        \hline
        B &0   &$0.10$ &0.62029 &0.61271 &0.99573 &0.99574\\ 
        E &10  &$0.09$ &0.86103 &0.85875 &0.99274&0.99274\\ 
        H &40  &$0.06$ &1.35036 &1.34378 &0.98960&0.98960\\ 
        K &100 &$0.04$ &2.02138 &2.01382 &0.98541&0.98541\\ 
    \end{tabular}
\end{ruledtabular}
\end{table}

In order to compare the frequencies of the scalar clouds resulting from the collapse of
boson stars with the known frequencies of quasi-bound states around Schwarzschild 
black holes, we consider next an scenario in which initially the black hole is already 
formed and has the same mass than that formed after the collapse of a boson star.
This initial Schwarzschild black hole is surrounded by a Gaussian spherical shell of 
scalar field which is evolved maintaining the background metric fixed. In this setup 
we find that after a short initial transient the field settles down into a long-lived mode 
akin to the ones showed in Figs.~\ref{fig:B_p} and~\ref{fig:H_p}. In order to 
characterize this field we Fourier-transform its amplitude to obtain the oscillation 
frequencies $\omega_{\text{qb}}^{(2)}$. The results are shown in the last column of   
Table~\ref{tab:dades2}. The excellent agreement between the frequencies 
$\omega_{\text{qb}}^{(1)}$ computed after black hole formation and  the 
frequencies $\omega_{\text{qb}}^{(2)}$ computed from the Gaussian pulse, 
is a clear indication that the configurations formed after the collapse of boson 
stars are indeed nonlinear quasi-bound states. 

Finally, in order to study the effect of $\Lambda$ on the frequencies and on the 
time decay of the quasi-bound states, we perform the same scattering experiment but 
keeping the mass $M_{\text{BH}}$ fixed. We find that for sufficiently long times $t\sim 
10^5 M_{\text{BH}}$ the effect of $\Lambda$ on the scalar field becomes 
negligible. This result is expected because the field decays exponentially and 
the dominating term is the the scalar-field mass. Therefore, despite the presence 
of nonlinear terms in the potential, the frequencies of all quasi-bound states will 
eventually tend to that of the quasi-bound state with $\Lambda=0$ (for the same 
black hole mass). The timescale to reach that situation depends on the value of 
$\Lambda$ because the frequency is different for each value of the coupling constant 
(both, the real and imaginary parts). Note that if we rescale the frequencies reported in 
Table~\ref{tab:dades2} with the BH mass, they do not coincide.

\section{Conclusions} \label{sec:conclusions}

We have presented a new numerical study of the Einstein-Klein-Gordon system in 
spherical symmetry. In particular we have discussed numerical relativity 
simulations of a large number of initial models of boson stars, both stable and 
unstable, and which incorporate a self-interaction potential with a quartic 
term.  Self-interaction provides extra pressure support against gravitational 
collapse, increasing the range of possible maximum masses of boson stars, 
allowing to encompass models with possibly larger astrophysical significance. We 
have revisited the stability of the initial solutions for different values of 
the self-interaction coupling constant $\Lambda$, as large as $\Lambda=100$, not 
previously considered in the literature (to the best of our knowledge). Our 
simulations have shown that the three different outcomes for unstable models, 
namely migration to the stable branch, total dispersion, and collapse to a black 
hole, reported before for the $\Lambda=0$ (mini-)boson star 
case~\cite{Seidel:1990jh,Balakrishna:1997ej,Guzman:2004jw}, are also present for 
self-interacting boson stars. We have focused our investigation on a subset of 
collapsing models, studying the effects the self-interaction potential may have in the 
presence of quasi-bound states. The existence of such long-lived quasi-bound 
states around black holes is supported by increasing numerical evidence, both 
based on perturbative calculations as on fully numerical relativity~\cite{Witek:2012tr,Burt:2011pv,Barranco:2012qs,Barranco:2013rua, sanchis2015quasistationary2,sanchis2015quasistationary,sanchis2016}. Moreover, 
they have been confirmed in a variety of physical systems, including both static 
and accreting black holes, and collapsing fermionic stars. In this work we have 
revisited this issue in the context of gravitationally unstable boson stars 
leading to black hole formation. We have found that for black-hole-forming 
models, a scalar field remnant can indeed be found outside the black hole 
horizon, oscillating at a different frequency than that of the original boson 
star. This result is in good agreement with recent spherically symmetric 
simulations of unstable Proca stars collapsing to black 
holes~\cite{sanchis2017numerical}.

\section*{Acknowledgements}

We thank Carlos Herdeiro and Jo\~ao G. Rosa for useful discussions and comments on the
manuscript. Work supported by the Spanish MINECO (AYA2015-66899-C2-1-P), by the 
Generalitat Valenciana (PROMETEOII-2014-069, ACIF/2015/216), by the CONACYT 
Network Project 280908 ``Agujeros Negros y Ondas Gravitatorias'', and by 
DGAPA-UNAM through grant IA103616. The computations have been performed at the 
Servei d'Inform\`atica de la Universitat de Val\`encia.


\bibliography{biblio}

\end{document}